\documentclass[%
 reprint,
 amsmath,amssymb,
 aps,
]{revtex4-2}

\usepackage{xcolor}
\usepackage{graphicx} 
\usepackage{braket}
\usepackage{enumitem}
\usepackage{nicefrac}
\usepackage{amssymb}
\usepackage{multirow}
\usepackage{soul}
\sethlcolor{yellow!50}

\soulregister\cite7
\soulregister\ref7
\soulregister\footnote7
\soulregister\textit7

\usepackage{amsmath}
\usepackage{array}
\usepackage{amsfonts}
\usepackage{bbm}
\usepackage{arydshln}
\usepackage{tikz}
\usepackage{qcircuit}
\usepackage{hyperref}
\usepackage{mathtools}
\usepackage[english]{babel}
\usetikzlibrary{decorations.pathmorphing}
\usetikzlibrary{shapes.geometric}
\newtheorem{lemma}{Lemma}

\newcommand{\be}{\begin{equation}}
\newcommand{\ee}{\end{equation}}
\newcommand{\equat}[1]{equation (\ref{eq:#1})}

\newcommand{\xv}{{\bf x}}

\newcommand{\Ex}{\mathbb{E}}

\newcommand{\Xparticle}{{\cal X}}
\newcommand{\Yparticle}{{\cal Y}}

\newcommand{\E}{{\cal E}} 
\newcommand{\Power}{{\cal P}} 
\newcommand{\Nzero}{{\cal N}_0} 
\newcommand{\Dp}{\Delta P_t}
\newcommand{\Amat}{{\bf A}}

\newcommand{\Cmat}{{\bf C}}
\newcommand{\Dmat}{{\bf D}}

\newcommand{\Kmat}{{\bf K}}
\newcommand{\Lmat}{{\bf L}}
\newcommand{\OImat}{{\mbox{\boldmath ${\cal O}$}}}
\newcommand{\Rmat}{{\bf R}}

\newcommand{\X}{\vec X}
\newcommand{\Y}{\vec Y}

\newcommand{\blankout}[1]{{}}

\newcommand{\equata}[2]{equations (\ref{eq:#1}) and (\ref{eq:#2})}

\newcommand{\Zbar}{{\overline{Z}}}

\newcommand{\Xbar}{{\overline{X}}}

\newcommand{\FlowR}{{\cal R}}

\begin{document}

\preprint{APS/123-QED}

\title{Limits of Information Flow Between Classically Interacting Particles}

\author{
Miles Miller-Dickson$^{1}$ and Christopher Rose$^{1}$}

\affiliation{$^{1}$School of Engineering, Brown University, Providence, RI, 02906}

\begin{abstract}
Pinning down a precise understanding of ``information flow'' within physical interactions remains a central concern to fields like stochastic thermodynamics and quantum information science. In both spheres a careful accounting of bits (or qubits) enables a deeper understanding of the physical nature of information. In this work we propose a measure of information flow as a saddle-point solution of the mutual information. This approach places a lower bound on the channel capacity between a particle and an interacting environment. The measure is given by ${\Power}_0/2{\cal E}_0$ in nats/sec, with ${\Power}_0$ the average power flux between the particle and its environment, and $\E_0$ the initial average energy of the particle, all computed in a frame where the particle has zero \textit{average momentum}. We use a communication theory lens to suggest an associated channel analogy, in which this bound is interpreted as a signal-to-noise ratio. We find that this measure can also quantify early-time information flow for a particle interacting with a thermal bath. 
\end{abstract}

\maketitle

\section{Introduction}
Incrementally, thought experiments like Maxwell's demon have led to increasingly abstract notions of information exchange between physical
constituents, such as cogs, levers, particles, etc. \cite{Smoluchowski1912,Feynman1963,Szilard1929} This abstracted information flow has partly culminated in the modern formulation of stochastic thermodynamics and the idea of the physics of information, in which information is treated on more equal footing with other physical entities such as energy. \cite{Parrando2015} 

Today, information flow in physical systems is more than ever an experimental question as much as a theoretical one. Yet despite a sense that information does indeed \textit{flow} between interacting particles, it has not been broadly established, outside of thermal equilibrium contexts, how much information is exchanged between particles precisely, say in nats/sec. Meanwhile, we are experiencing a period of unprecedented control over microscopically fluctuating quantum and classical systems in which to explore fundamental bounds on information transfer. Examples include experimental verifications of Landauer's principle \cite{Landauer-test-1,Landauer-test-2} and verifications of Jarzynski/Crooks-type fluctuation relations, which have information theoretic ramifications. \cite{Jarzynski-test-1,Jarzynski-test-2} Moreover, in quantum systems careful accounting and control over information flow, or more specifically information \textit{loss}, will be necessary to push beyond the noisy intermediate-scale quantum era (NISQ). \cite{Preskill2018} 

Our paper considers the question of information exchange along two new avenues. One is that while previous approaches have largely derived statistical properties from equilibrium or steady state assumptions in one form or another, in this work we use a communication theory lens to consider information flows in fully out-of-equilibrium settings. Secondly, we consider information flow purely from the perspective of \textit{interacting particles}, as characterized by the couplings between degrees of freedom, whereas prior work has largely considered the information content that a laboratory experimenter might have regarding a system under study (even if that experimenter is a Maxwellian demon). In contrast, particles are ``naive'' entities, whose only memory storage to speak of is its current state in phase space, and whose only information processing capability is to respond to forces from its environment. 

The consideration of information flow between microscopic particles has not received as much attention in the literature on the physics of information, primarily due to the difficulty of defining what is meant by information flow in these settings. Information theory is based on probability measures over states, representing either uncertainty that one entity has with respect to another, or the fraction of time spent in each state, depending on your viewpoint. In any case, it is difficult to justify the assignment of one probability measure versus another to a particle state without being able to make use of some kind of asymptotic equipartition property or ergodic hypothesis that is afforded by equilibrium assumptions.

In this work, however, we sidestep this difficulty by considering the full range of priors over particle and environment states. We determine an upper bound on the total information flow from the environment to the particle in a maximally noisy setting, subject to energy and power flow constraints. More concretely, we consider a deterministic map of the form $X' = f_t(X,Y)$, such as Hamiltonian flow, that takes in random particle ($X$) and environment states ($Y$) at time $0$, and returns the particle state at time $t>0$ ($X'$). Then one can ask about the mutual information between $Y$ and $X'$, the information exchange between the environment and the particle after a time $t$. One can regard $X'$ as a \textit{compressed} version of the previous particle and environment states, for which we can determine the extent that $X'$ encodes the interaction from $Y$. In this context, we thus treat $Y$ as an input signal that updates a previous particle state $X$ to the value $X'$, which is the output response. Hence $X$ is treated as a noise variable which interferes with the mapping from $Y\to X'$. In the small $t$ limit, we linearize $f_t$ and construct an additive channel, for which we compute the maximum mutual information $I(X';Y)$ for ``signal'' and ``noise'' power constraints. When $X,Y,X'$ are momenta, we will be able to interpret these constraints as true power flow and energy conditions for the particle in the \textit{zero average momentum frame}, as discussed below.

Our result represents a \textit{saddle-point} in the mutual information between the environment and particle states, in a classical, non-relativistic setting. That is, the mutual information $I(X';Y)$ is a \textit{concave} functional of the signal distribution $p_Y$, but is a \textit{convex} functional in the noise distribution $p_{X'|Y}$. There exists a saddle-point in the space of distributions for which the mutual information gives the \textit{lowest} channel capacity (i.e. lowest maximum information flow) across the space of possible channel distributions $p_{X'|Y}$. As a saddle-point, this solution also represents the maximal information flow in the noisiest environment, across all input distributions $p_Y$. The saddle-point therefore provides a meaningful measure of the information flow, while also relating information theoretic quantities to energy scales in a new way. 

The mutual information is a logarithmic measure of the reduction in the space of possibilities for one variable when given (conditioned on) the other. We identify it here with information flow as it represents the extent that a particle state \textit{distinguishes} between environment states. That is, if as a result of an interaction, a particle state encodes its environment---and before this period the particle and environment were otherwise independent---then knowledge of the particle state reduces the space of environmental states, and the extent of this reduction can be captured by the mutual information. \footnote{We assume that the physical laws governing the interaction are fixed and given, so that the information flow is ultimately the result of that structure.} An essential ingredient in this definition is the \textit{size} of the space over which particle and environment states can effectively vary, which will be characterized here by the variance of the priors over the states. We derive these priors, which turn out to be Gaussian, from assumptions of bounded \textit{energy} and \textit{power flow} between the particle and environment, placing limits on this variability. These bounds on energy and power allow one to discuss information flow with respect to energy and time scales. It will turn out that constraining the average kinetic energy of a particle to a value $\E_0$, and the average power flow from the environment to ${\Power}_0$, yields an information flow saddle-point given by the ratio, ${\Power}_0\big/2\E_0$, in nats/sec. This ratio gives the maximal information flow \textit{rate} in a setting of maximal noise.

A related method for measuring information flow is the transfer entropy (TE). \cite{Schreiber2000} This is a conditional mutual information between an environment's history and a system's future, conditioned on the history of the system. It gives the additional predictive power that the environment's history has on the future state of the system, beyond what could be predicted from the system's history alone. In conditioning on the history of the system, one thus isolates only the information content that originates with the environment. Our approach is similar in that it quantifies how much the past of the environment can be used to predict the future of a particle state, but for the simple systems we consider here, conditioning will lead to infinite TE. Instead, we isolate the effect of the environment by considering settings in which the environment and particle are initially \textit{independent}, and for the systems we consider here the time-delayed mutual information we compute yields a finite measure of information flow. \footnote{Under the \textit{noisiest} setting, given the constraints, we ask what is the most information that can be transmitted from environment to particle. This question is more meaningful than asking, for instance, what is the most information that can be transmitted in the \textit{least} noisy environment, since that would correspond to zero noise (no variance in the particle prior) and the particle state would exactly encode effect of the environment. The information exchanged would be infinite in this case (for a continuous state space), which is not particularly useful. This highlights the difficulty in defining information flow in a physical setting. The information exchanged depends on the priors over the particle and the environment, but the priors over such variables are context-dependent: The particle could be constrained to be within a box, or in thermal equilibrium at some temperature, etc. Our approach, therefore, is to consider a bound on the mutual information over the space of \textit{all} priors such that the mutual information is \textit{maximized} with respect to priors on the environment and \textit{minimized} with respect to priors on the particle state (a maximal signal in a maximal noise setting). We will see that this quantity is finite and hence places a lower bound on the maximum (minimax) information flow between a particle and its environment.} 

In section~\ref{sect:II} we motivate our main result using a reduced model for the the state of a particle. This simple example captures the main idea and will give an identical formula as in the full state space case. In section~\ref{sect:III}, we consider the full state space and consider a short time limit for the interaction, which is appropriate for discussing information \textit{flow}, i.e. information exchange on very short timescales. In section~\ref{sect:IV} we investigate a springlike interaction model between two 1-dimensional particles, and
compute the corresponding mutual information as a function of the coupling constant. In section~\ref{sect:V} we conclude with a brief discussion of the implications of these results and directions for future study.

\section{A motivating example}
\label{sect:II}
\begin{figure}[h!]
\begin{tabular}{lr}
{\bf (a)} & 
\includegraphics[width=2.75in]{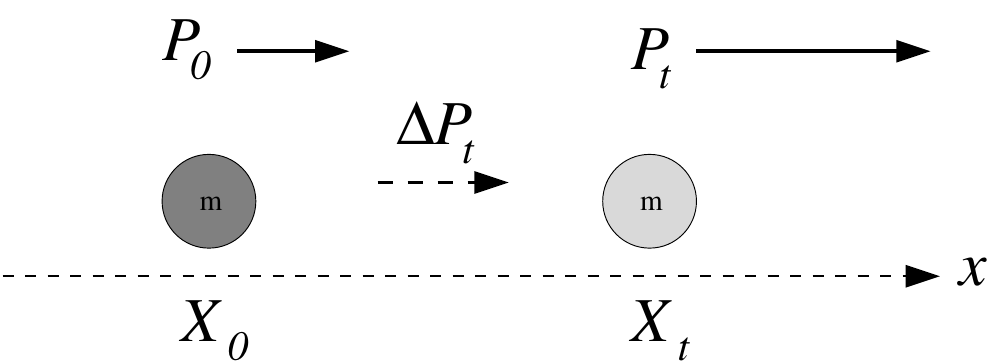}\\
$\mbox{ }$ & $\mbox{ }$\\
{\bf (b)} & 
\includegraphics[width=3.0in]{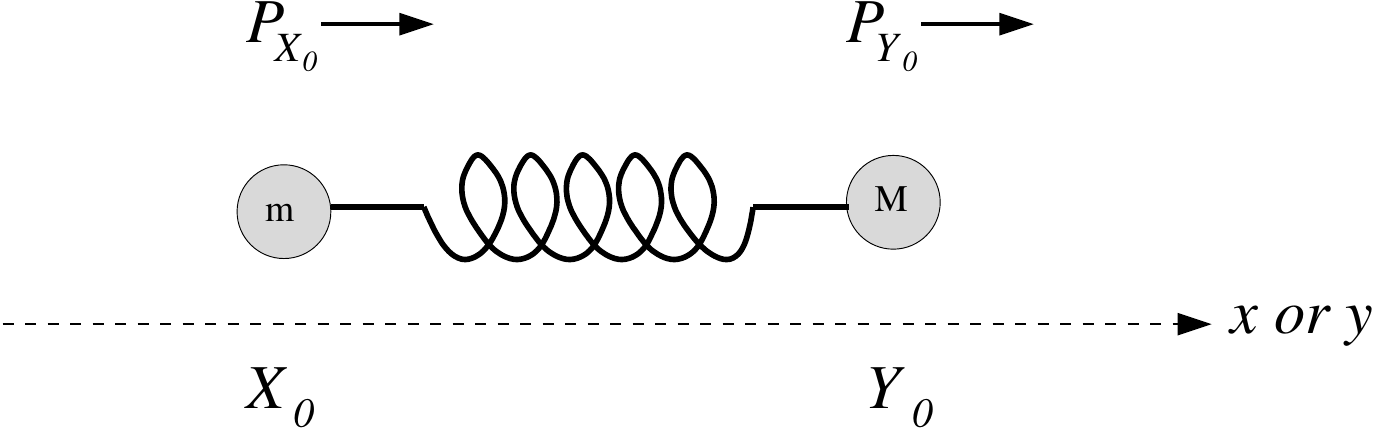}\\
 \end{tabular}
    \caption{(a) A classical particle in one dimension under the influence of a random environment (analyzed in sections \ref{sect:II} and \ref{sect:III}). $X_0$ and $P_0$ are the position and momentum respectively at time $0$, and $X_t$ and $P_t$ give the position and momentum at $t$.  $\Delta P_t$ is the change in momentum. (b) Two particles interacting with a springlike interaction (analyzed in section \ref{sect:IV}).}
    \label{fig:diagram}
\end{figure}

Consider a classical, non-relativistic particle on the real axis, under the influence of some interaction with an environment, which has been turned on at time $0$ (see FIGURE~\ref{fig:diagram}(a) below).
The particle state at time $0$ is given by the pair of random variables $(X_0,P_0)$, of position and momentum, although we can illustrate the essential idea by just considering the momentum variable $P_0\in\mathbb{R}$ for now. We will ask how the momentum of the particle at some \textit{fixed} time $t>0$ encodes the interaction with the environment, as measured by the mutual information. That is, the momentum after a time $t$ can be written as,
\begin{align}
    \label{eq:basic-channel}
        P_t &=P_0 + \Delta P_t
\end{align}
in which $\Delta P_t$ encompasses all effects from the environment in that time. An example is Brownian motion for which $\Delta P_t =\sum_{t'\le t}\delta_{t'}$ for a series of impulses $\delta_{t'}$. An interaction of this type is commonly modeled in stochastic thermodynamics by a Langevin-type equation. \cite{Rastegar2012,Seifert2012} In that context, one typically splits this effects of the environment into drift and noise terms, however, our interest here is the total influence the environment has on the particle state. Thus, we lump all effects into the singular variable $\Delta P_t$.

The mutual information between the particle state at time $t$ and the effect of the environment is given by, 
\be
    \label{eq:mutInfo1}
    I(P_t;\Delta P_t) = h(\Delta P_t) - h(\Delta P_t|P_t) \ge 0
\ee
where $h(X) = -\int p_X(x) \log p_X(x) dx$ is the differential entropy. \cite{CoverAndThomas} The mutual information is a logarithmic measure of the number of states for $\Delta P_t$ that can be distinguished by states of the momentum $P_t$ (and vice versa). In the case that $\Delta P_t$ and $P_0$ are assumed independent, so that $I(P_0;\Delta P_t)=0$, the mutual information $I(P_t;\Delta P_t)\ge 0$ gives a measure of the information the particle's momentum gains in a time $t$ from the environment. We therefore assume that $\Delta P_t$ is independent of the initial momentum $P_0$. \footnote{As a point of comparison, the transfer entropy for this example is \textit{infinite}. The TE can be written as $T_{F\to P}=I(P_t;\Dp|P_0)$, which is the mutual information version of equation (4) in \cite{Schreiber2000} with $k=l=1$ and with $n$ and $n+1$ corresponding to the time points $0$ and $t$, respectively. However, in this toy example $I(P_t;\Dp|P_0)=I(P_t;\Delta P_t)+I(P_0;\Delta P_t|P_t)$, since $I(P_0;\Dp)=0$, but the latter quantity explodes, $I(P_0;\Dp|P_t)=\infty$, whereas the former quantity is the finite value we compute, $I(P_t;\Dp)$. The latter quantity, $I(P_0;\Dp|P_t)$, explodes because $P_0$ and $P_t$ together determine the change in the momentum $\Dp$ exactly, and specifying an arbitrary point on the real line, $\Dp\in\mathbb{R}$, requires infinite information. Thus the transfer entropy is not as useful in this context, that is unless one allows for a case like $k=0$ in \cite{Schreiber2000}, for which one could say our approach is a \textit{zeroth order} transfer entropy, but that is just the time-lagged mutual information without conditioning.}   

Without constraints the mutual information is unbounded from above, and so we set two energy conditions on the variables $P_0$ and $\Delta P_t$. First, we fix the initial average energy of the particle:
\be
    \label{eq:energy-bound}
    \Ex\left[\frac{P_0^2}{2m}\right]=\E_0
\ee
for some given energy scale $\E_0$, where the expectation $\Ex[.]$ is taken over the prior distribution for particle states $P_0$. We've assumed here that our particle has mass $m$.

Second, we assume that the average energy gained in a frame where the average momentum is \textit{zero} is constrained by,
\be
    \label{eq:power-bound}
    \Ex\left[\frac{(\Delta P_t)^2}{2m}\right]= \E_t
\ee
for some energy $\E_t$. In such a frame, the average energy given to the particle in a time $t$ is,
\begin{align}
\begin{split}
    \Ex[\Delta E]&=\Ex\left[\frac{(P_0+\Delta P_t)^2}{2m}-\frac{P_0^2}{2m}\right] \\
    &=\Ex\left[\frac{P_0\Delta P_t}{m}\right]+\Ex\left[\frac{(\Delta P_t)^2}{2m}\right]
\end{split}    
\end{align}
and if $\Ex[P_0]=0$, and $P_0$ is assumed independent of $\Delta P_t$, then we see that $\Ex[(\Delta P_t)^2/2m]\ge0$ gives the average energy gain in this frame. An even stronger condition is that the frame is co-moving with the particle so that $P_0=0$, i.e. no variance in the initial momentum, though the vanishing of the \textit{average} initial momentum suffices for our purposes.

These conditions along with \equat{basic-channel} can be treated as an additive noise channel with variance constraints. We treat the previous momentum state $P_0$ as a \textit{noise} term, and the term $\Delta P_t$ as an \textit{input signal} from the environment. $P_0$ is a noise term since we're interested in the communication channel between the environment $\Delta P_t$ and the updated particle state $P_t$. The $P_0$ term introduces additional variance in the output $P_t=P_0+\Delta P_t$ which interferes with the mapping from $\Delta P_t\to P_t$. 

Given the constraints in \equata{energy-bound}{power-bound}, jointly Gaussian $P_0$ and $\Delta P_t$ yield a saddle-point for the mutual information in \equat{mutInfo1}. \cite{Ihara78, McEliece1983, BordenMasonMcEliece85} More concretely, it can be shown that,
\begin{equation}
\label{eq:saddle-point}
\begin{gathered}
     I(P_0^*+\Dp;\Dp)  \\
     \le \\
     I(P_0^*+\Dp^*;\Dp^*) \\
     \le \\
     I(P_0+\Dp^*;\Dp^*)  
\end{gathered}
\end{equation}
for $P_0^*$ and $\Dp^*$ \textit{zero-mean} Gaussian random variables satisfying the constraints: $P_0^*\sim{\cal N}(0,2m\E_0)$ and $\Dp^*\sim{\cal N}(0,2m\E_t)$.  While this result is well-known in the communication theory community \cite{CoverAndThomas,Ihara78, BordenMasonMcEliece85, McEliece1983} we provide a proof of the $1$-dimensional case in the appendix, \ref{app:minimax}, for clarity.  

Under these constraints the saddle-point mutual information above is easily computed,
\begin{align}
    \begin{split}
    \label{eq:SNR}
    I^*_t \equiv I(P_0^*+\Dp^*;\Dp^*) &= \frac{1}{2}\log\left(1+\frac{\E_t}{\E_0}\right)\\
    &\approx\frac{1}{2}\left(\frac{\E_t}{\E_0}\right)
    \end{split}
\end{align} 
and the approximation holds when the ratio of energies is sufficiently small, $\E_t/\E_0\ll 1$. This can occur, for instance, when $\E_t = {\Power}_0t$ for some average power $\Power_0$ and the time interval $t$ is such that $t\ll\E_0/{\Power}_0$.

In this case, we can define the corresponding information flow \textit{rate} $\FlowR \equiv \lim_{t\to 0}(I^*_t\big/t)$:
\be
    \label{eq:SNR1}
    \FlowR_{env \to particle}=\frac{1}{2}\left(\frac{{\Power}_0}{\E_0}\right)
\ee
yielding a signal-to-noise ratio for interaction with the environment. This result will turn out to be the same as in the full state space and represents the main contribution of this work.

\section{The Full State Space}
\label{sect:III}
We now extend this result to the full state space of a one-dimensional particle comprised of the position-momentum pair, $(X_0,P_0)$. The evolution of the particle state in a small time $t$ under the influence of an environment is given by, 
\begin{align}
\label{eq:Xt}
\begin{split}
     X_t &= X_0 + \frac{1}{m} \left(\frac{P_0 + P_t}{2}\right)t\\
 P_t &= P_0 + \Dp 
\end{split}
\end{align}
and writing the average momentum as $(P_0+P_t)/2=P_0+(1/2)\Dp$, we isolate the signal and the noise pieces,
\be
\label{eq:XF_channel}
\X_t = \underbrace{\left(\begin{array}{cc}
    1 & t/m \\
    0 & 1
\end{array}\right)\X_0}_\text{noise $\equiv\vec{W}_t$} + 
\underbrace{\left(\begin{array}{c}
      t\big/2m  \\
      1
\end{array}\right)\Dp}_\text{signal $\equiv\vec{F}_t$}
\ee
where we treat the initial conditions, 
\be
\X_0\equiv
\left ( 
\begin{array}{c}
X_0\\
P_0
\end{array}
\right )
\ee
as ``noise,'' labeled by $\vec{W}_t$, and we regard $\Dp$ as a ``signal,'' labeled by $\vec{F}_t$. Together, $\vec{W}_t$ and $\vec{F}_t$ form an additive channel with output $\X_t=(X_t,P_t)^\top$. We assume independence of the variables $X_0,P_0$ and $\Dp$.

In addition to the energy constraints in equations \ref{eq:energy-bound} and \ref{eq:power-bound} above, we introduce a bound on the position variable $X_0$, for which we assume, without loss of generality, that $\Ex[X_0]=0$, and that,
\be
    \Ex[X_0^2]<\infty
\ee
so that the position of the particle is confined.

As before it can be shown that under such constraints on the variances, the saddle-point value for the mutual information,
\be
    I(\X_t;\Dp)=h(X_t,P_t) - h(X_t,P_t|\Dp) 
\ee
is achieved for jointly and independent zero-mean Gaussian variables, $\Dp$ and $\X_0=(X_0,P_0)$, as shown in \cite{Diggavi2001}.

Set the covariance of $\X_0$ to be $\Kmat_0$. The covariance of $\X_t$ is $\Kmat_t\equiv\Kmat_{W_t}+\Kmat_{F_t}$, the sum of the covariances of the noise and the input signal, respectively, due to the independence of $\X_0$ and $\Dp$. Whereas, the conditional entropy is given by $h(X_t,P_t|\Dp)=h(\vec{W}_t)$, since the channel is additive and the noise is independent of the signal. Using the entropy formula for an $n$-dimensional Gaussian, $\frac{1}{2}\log((2\pi e)^n|\Kmat|)$, with covariance $\Kmat$, the mutual information can be written, 
\be 
\label{eq:XF_fullMutInfo}
I(\X_t;\Dp)=\frac{1}{2}\log\left(\frac{|\Kmat_{W_t}+\Kmat_{F_t}|}{|\Kmat_{W_t}|}\right)
\ee
where $|\cdot|$ indicates the determinant. 

The covariances of the noise and the input signal are given explicitly by,
\begin{align}
\begin{split}
\Kmat_{W_t} &\equiv \left(\begin{array}{cc}
    1 & t/m \\
    0 & 1
\end{array}\right)\Kmat_0\left(\begin{array}{cc}
    1 & 0 \\
    t/m & 1
\end{array}\right)
\\
\Kmat_{F_t} &\equiv \Ex[(\Dp)^2]\left(\begin{array}{cc}
    t^2/4m^2 & t/2m \\
    t/2m & 1
\end{array}\right)
\end{split}
\end{align}
with $\Kmat_0$ the covariance of the initial conditions, $\X_0$. Assuming the initial position and momentum are independent:
\be
\label{eq:cov1}
\Kmat_0=\left(\begin{array}{cc}
    \Ex[X_0^2] & 0 \\
    0 & \Ex[P_0^2]
\end{array}\right)
\ee
\equat{XF_fullMutInfo} can written out explicitly and expanded to leading order in $t$,
\begin{align}\begin{split}
\label{eq:XF_approxMutInfo}
&I(\X_t;\Dp) \\
&=\displaystyle\frac{1}{2}\log\left(1+\frac{\Ex[(\Dp)^2]}{\Ex[P_0^2]}\left(1+\frac{\Ex[P_0^2]t^2}{4m^2\Ex[X_0^2]}\right)\right) \\
&\displaystyle \approx\frac{\Ex[(\Dp)^2]}{2\Ex[P_0^2]}\\
&=\frac{1}{2}\frac{{\Power}_0}{\E_0}t
\end{split}
\end{align}
where in the final step we assumed $\Ex[(\Dp)^2] = 2m{\Power}_0 t$, and used \equat{energy-bound} to write $\Ex[P_0^2]=2m\E_0$. This is the same result we found in \equat{SNR}. The dependence on the position variance $\Ex[X_0^2]$ does not contribute to leading order since $X_0$ enters into \equat{Xt} without any time factors $t$ (compare to $P_0$ and $\Dp$ in \equat{SNR} which are both scaled by $t$ in the $X_t$ equation). In taking the ratio $|\Kmat_{W_t}+\Kmat_{F_t}|/|\Kmat_{W_t}|$, the determinant $|\Kmat_{W_t}|=\Ex[X_0^2]\Ex[P_0^2]$ cancels the $\Ex[X_0^2]$ dependence, leaving the ratio $\Ex[(\Dp)^2]/2\Ex[P_0^2]$ at leading order.

This result shows that more information is stored in the state $\X_t$ about its environment when the variance of the momentum exchange $\Dp$ is relatively large, or when the initial momentum variance is relatively small. A larger variance in $\Dp$ enables a wider space of environments that can be communicated to the particle state. And a narrower variance in $P_0$ allows for a sharper estimate of the forces from the environment, i.e. of $\Dp\big/t$. In terms of energy, an interaction from the environment communicates more to the particle when its energy is large compared to the particle's initial energy.

Furthermore, we observe that heavier particles, which have more momentum, all things being equal, will have higher momentum variance, and will thus store \textit{less information} in their state about their environment in time $t$. Larger masses effectively reduce signal strength by lowering the acceleration from environmental driving, diminishing the distinction between $\X_0$ and $\X_t$ in a time $t$. Thus, larger information exchanges are predicated upon improved distinguishability between states. This occurs when the range of driving by the environment is greater (or the mass is smaller), or from increased precision in starting states, reducing ambiguity about which starting states lead to which final states. 

\section{Two Particles With A Springlike Coupling}
\label{sect:IV}
We now consider a model in which the mutual information can be computed for arbitrary times $t$: An interaction between two one-dimensional
particles that exchange energy through a quadratic (springlike) potential, as shown in FIGURE~\ref{fig:diagram}(b). This model has also been considered in the stochastic thermodynamics literature, for instance in \cite{Herpich2020}, although we do not make use of thermal sources in this work. Call the particles $\Xparticle$ and $\Yparticle$, with masses $m$ and $M$, respectively. We imagine that the particles pass through each other without effect. The Hamiltonian
for such a system can be written as,
\be
    H = \frac{P_X^2}{2m} + \frac{P_Y^2}{2M} + \frac{1}{2}k(X - Y)^2 
\ee
where $k$ is the spring constant, setting the coupling strength between the particles. $X(t)$ and $Y(t)$ are the positions of the $\Xparticle$ and $\Yparticle$ particles respectively. The dynamics of this system are linear and can be written as,
\be
\dot{\xv}_t
=
\Amat \xv_t.
\ee
where,
\be
\xv_t\equiv
\left (
\begin{array}{c}
X(t)\\
P_X(t)/m\omega\\
Y(t)\\
P_Y(t)/M\omega
\end{array}
\right) \equiv \left(\begin{array}{c}
     \X_t \\ 
     \Y_t
\end{array}\right)
\ee
\be
\X_t
\equiv
\left (
\begin{array}{cccc}
1 & 0 & 0 & 0 \\
0 & 1 & 0 & 0 
\end{array}
\right ) \xv_t
=
\Cmat \xv_t
\ee
and
\be
\Y_t
\equiv
 \left (
\begin{array}{cccc}
0 & 0 & 1 & 0 \\
0 & 0 & 0 & 1 
\end{array}
\right ) \xv_t
=
\Dmat \xv_t
\ee
where we've used the masses and the natural frequency $\omega=\sqrt{k(m+M)/mM}$ to build position-like vectors, representing the states of each particle at time $t$. Note that the mutual information is invariant to scaling the state vectors in an invertible manner like we've done here. $\Amat$ is given by,
\be
\label{eq:Amat}
\Amat = 
\left (
\begin{array}{cccc}
0 & \omega & 0 & 0 \\
-k/m\omega & 0 & k/m\omega & 0 \\
0 & 0 & 0 & \omega \\
k/M\omega & 0 & -k/M\omega &  0 
\end{array}
\right ).
\ee
The solution can be written as $\xv_t = \Phi(t) \xv_0$, where $\Phi(t) = e^{\Amat t}$, given explicitly by,
\be
\label{eq:Phi}
\Phi(t)=\left (
\begin{array}{cc}
    \Lmat_t & \Rmat_t \\
    \tilde\Rmat_t & \tilde\Lmat_t 
\end{array} \right )
\ee
where, 
\begin{align*}
    \Lmat_t&\equiv\frac{m}{m+M}\left(
    \begin{array}{cc}
    1+\frac{M}{m}\cos\omega t & \omega t + \frac{M}{m}\sin\omega t \\[0.1cm]
    -\frac{M}{m}\sin\omega t & 1+\frac{M}{m}\cos\omega t 
    \end{array}\right), \\[0.1cm]
    \Rmat_t&\equiv\frac{M}{m+M}\left(
    \begin{array}{cc}
    1-\cos\omega t & \omega t - \sin\omega t \\
    \sin\omega t & 1-\cos\omega t
    \end{array}\right)
\end{align*}
with $\tilde\Lmat_t$ and $\tilde\Rmat_t$ defined analogously but with $m$ and $M$ swapped. 

The dynamics for the $\Xparticle$ particle can be written as,
\be
\label{eq:xA_t}
\X_t = \Lmat_t \X_0 + \Rmat_t\Y_0
\ee

As in \equat{XF_channel}, we treat the first term as a noise term while the second contributes to the signal from the other particle. Two limits worth noting are when $M\to 0$ and $M\to\infty$ (with $m>0$ fixed). In the first case, there is effectively no mass to couple to, $\Rmat_t\to \left(\begin{array}{cc}
    0 & 0 \\
    0 & 0
\end{array}\right)$ and $\Lmat_t$ reduces to the case of constant uniform motion (this is equivalent to sending $k\to 0$, and the mutual information vanishes). In the latter case, when the second mass is infinite, $\Lmat_t$ reduces to a rotation matrix (for $t\ll M/m\omega$) and the ratio $M/(m+M)\to 1$ in $\Rmat_t$, resulting in simple harmonic motion in a moving frame. We will consider the information flow for various mass ratios, $M\big/m$. 
\subsection{Mutual information $I(\X_t;\Y_0)$ for $t\ge 0$}
\label{sect:IVa}
We regard the $\Xparticle$ system as a probe that interacts with the $\Yparticle$ particle and
``measures'' it. We assume that the $\Xparticle$ and $\Yparticle$ systems are initially
\textit{independent} and that their interaction is switched on at time $0$. We ask how the $\Xparticle$
state (position and momentum) at a given time $t$ encodes the state of the $\Yparticle$ particle in the case of Gaussian priors. Here we relax the assumption that $t$ be small (compared to some interaction timescale) for improved generality---although we will subsequently consider the small $t$ limit for comparison, for which the saddle-point solution for the mutual information is provided by Gaussian priors for the same reasons as before. 

In this two-particle example, we can regard the
$\Yparticle$ system as the environment for the $\Xparticle$ particle, and we therefore elect to consider the mutual information between particle \textit{states}, rather than between a state and a change in momentum---we will find in the small $t$ limit however that the result is the same as before. 

Note that the state $\Y_0$ cannot be fully determined by the state $\X_t$, at a specific
time $t$. That said, the {\em trajectory}
$\{\X_\tau\}_{\tau\in[0,t]}$ {\em can} be used to determine $\Y_0$ since the observability matrix has full column rank, as we discuss in the appendix, \ref{app:observability}. \cite{ControlTheory2011} Nonetheless, our information-theoretic framing shows that $\X_t$ can provide {\em some} information about $\Y_0$---and at specific times can specify a component of $\Y_0$ exactly.

The information stored in $\X_t$ about $\Y_0$ is the mutual information,
\be
I(\X_t;\Y_0) = h(\X_t) - h(\X_t|\Y_0).
\ee
We assume $\X_0$ and $\Y_0$ are independent, and so the covariance of $\X_t$ is additive in the covariances of the variables $\vec{L}_t\equiv\Lmat_t\X_0$ and $\vec{R}_t\equiv\Rmat_t\Y_0$ (\equat{xA_t}), and furthermore we have that $h(\X_t|\Y_0)=h(\vec{L}_t)$, analogous to our motivating example. Thus we can write the mutual information as,
\be
\label{eq:XY_fullMutInfo}
I(\X_t;\Y_0)=\frac{1}{2}\log\left(\frac{|\Kmat_{L_t}+\Kmat_{R_t}|}{|\Kmat_{L_t}|}\right),
\ee
analogous to \equat{XF_fullMutInfo}, with covariances,
\be
\Kmat_{L_t}\equiv\Lmat_t\Kmat_{X_0}\Lmat_t^\top\mbox{  \text{and} }
\Kmat_{R_t}\equiv\Rmat_t\Kmat_{Y_0}\Rmat_t^\top,
\ee
with diagonal initial covariances $\Kmat_{X_0}$ and $\Kmat_{Y_0}$, as in \equat{cov1}.

\begin{figure}
    \centering
    \includegraphics[scale=0.55]{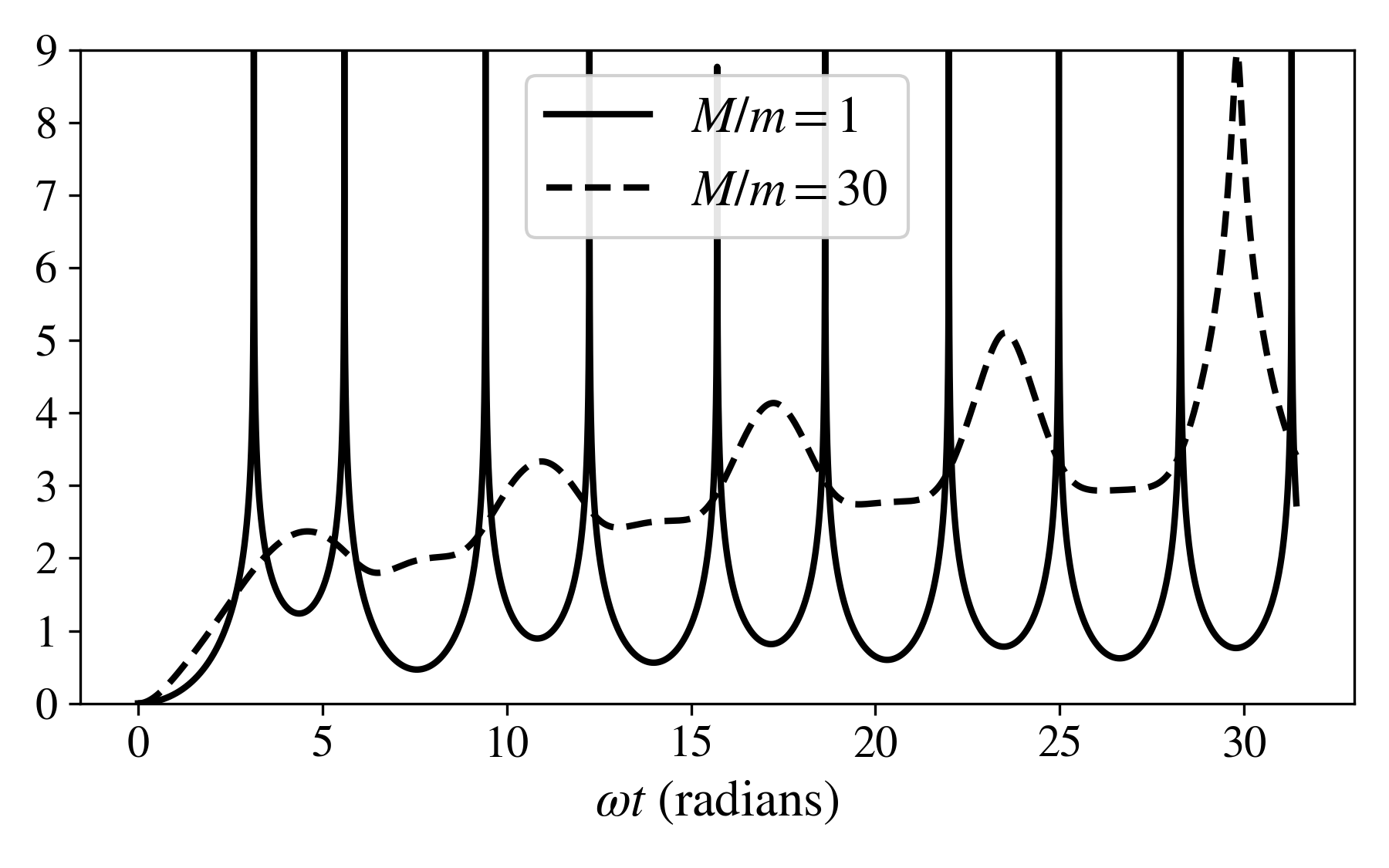}
    \caption{{\bf Mutual Information:} We plot the mutual information (in \textit{nats}) for the springlike model $I(\X_t;\Y_0)$ for two mass ratios. Parameters are set so that $\Kmat_{X_0}=\Kmat_{Y_0}=\left(\begin{array}{cc}
        1 & 0 \\
        0 & 1
    \end{array}\right)$. Spikes indicate divergence toward $+\infty$. 
    }
    \label{fig:mutualInfo}
\end{figure}

The mutual information is plotted in FIGURE \ref{fig:mutualInfo} for mass ratios $M/m=1$ and $M/m=30$. Most notable are the infinite spikes in the mutual information where the determinant $|\Kmat_{L_t}|=|\Lmat_t|^2|\Kmat_{X_0}|$ vanishes. The determinant of $\Lmat_t$ is given by,
\be
\label{eq:Lt_det}
|\Lmat_t| = \frac{m^2+M^2+2mM\cos{\omega t}}{(m+M)^2}+\frac{mM\omega t}{(m+M)^2} \sin{\omega t}
\ee
The first term is bounded between $(m-M)^2/(m+M)^2$ and $1=(m+M)^2/(m+M)^2$, while the second term has an envelope with a magnitude that grows linearly in $t$, eventually causing the determinant to oscillate across the x-axis. After that point, the determinant periodically crosses the axis leading to the spikes shown in the mutual information. 

At these spikes, $\X_t$ exactly encodes one of the two dimensions of the initial $\Yparticle$ state, $\Y_0$ (at every other spike the initial $\Yparticle$ velocity is exactly encoded in $\X_t$; at the remaining spikes a particular linear combination of the initial $\Yparticle$ position and velocity is encoded by $\X_t$). 

As the mass ratio $M/m$ increases the initial spike in the mutual information is pushed farther out in time. This can be seen from the slope of the linear envelope in the determinant of $\Lmat_t$ (\equat{Lt_det}), which scales like $m\omega/M$ for $M/m\gg1$ (for fixed $k$, $\omega\to\sqrt{k/m}$ as $M\to\infty$). In this limit the $\Xparticle$ particle cannot encode its environment as efficiently, and the growth of mutual information from $\Yparticle$ to $\Xparticle$ is more gradual (shown by the dashed curve in FIGURE \ref{fig:mutualInfo}). 

Conversely, we note that in the limit as $M\to 0$, the mutual information vanishes---$\Xparticle$ encodes nothing about $\Yparticle$, which was also demonstrated in \cite{Herpich2020}. When $M\to\infty$ then $|\Lmat_t|\to1$ and the mutual information grows logarithmically in time, reflecting the growth of the \textit{relative} precision that the $\Xparticle$ state has in approximating the $\Yparticle$ state. Evidently, information flows more readily when the masses are matched; near $M\approx m$ the mutual information spikes quickly.
\begin{figure}
    \centering
    \includegraphics[scale=0.5]{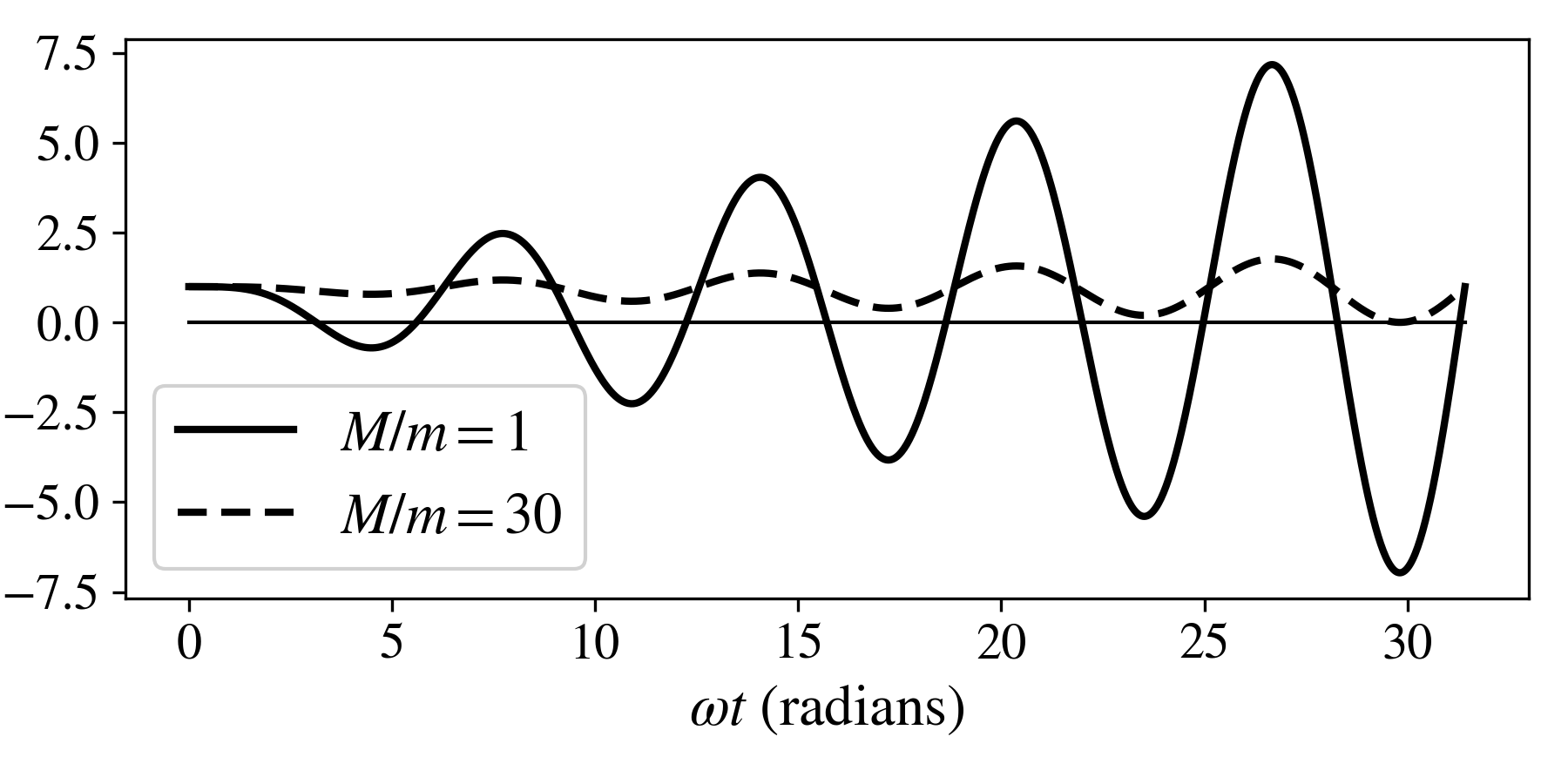}
    \caption{{\bf Determinant of $\Lmat_t$:} We plot $|\Lmat_t|=(m^2 + M^2 + 2mM\cos\omega t + mM\omega t\sin\omega t)/(m+M)^2$ as a function of $\omega t$ to show how the mass ratio, $M/m$, determines the first zero-crossing for the determinant, and thus the first spike for the mutual information (see FIGURE \ref{fig:mutualInfo}).}
    \label{fig:DetR}
\end{figure}

\subsection{Small $t$ limit: Info Rate from \texorpdfstring{$\Yparticle$}{LG} to \texorpdfstring{$\Xparticle$}{LG}}
We have considered the mutual information between the state $\X_t$ and the state $\Y_0$. We would like to compare this to our result in \equat{XF_approxMutInfo}, the information flow between a particle and the environment acting on it in a small time interval. We again make use of the constraints in \equat{energy-bound} and \equat{power-bound}, but now we specialize the latter to the case of the spring force, writing $\Dp = F_0 t$ for a small interval $t$, with $F_0\equiv -k(X_0-Y_0)$, the initial force on the $\Xparticle$ particle. Then, the power constraint in \equat{power-bound} translates to a constraint on the force,
\begin{align}
\label{eq:power-bound2}
    \begin{split}
        \Ex[F_0^2] &= k^2\Ex[(X_0-Y_0)^2]\\
        &=k^2(\Ex[X_0^2] + \Ex[Y_0^2]) \\ 
        &=\frac{2m{\Power}_0}{t}
    \end{split}
\end{align}
where we've assumed independence and the zero-mean condition to set $\Ex[X_0Y_0]=0$, and furthermore that $\Ex[(F_0t)^2]=2m{\Power}_0t$ as in equations \ref{eq:SNR1} and \ref{eq:XF_approxMutInfo}. This constraint assumes finite moments for $\Ex[X_0^2]<\infty$ and $\Ex[Y_0^2]<\infty$, for $t>0$, and that ${\Power}_0$ is finite. 

We assume that $\omega t\ll 1$. In this limit, $\Lmat_t$ and $\Rmat_t$ can be approximated to first order in $t$ by,
\be
\Lmat_t \approx
\left(
\begin{array}{cc}
1 & \omega t \\
-\frac{M\omega t}{m+M} & 1
\end{array}
\right),
\Rmat_t
\approx
\frac{M}{m+M}\left(
\begin{array}{cc}
0 & 0 \\
\omega t & 0
\end{array}
\right).
\ee
The mutual information to leading
order in $t$ is given by,
\be
\label{eq:XY_approxMutInfo}
I(\X_t;\Y_0)\approx\frac{1}{2}\log\left(1+\frac{k^2\Ex[Y_0^2]t^2}{\Ex[P_0^2]}\right)\approx\frac{k^2\Ex[Y_0^2]t^2}{2\Ex[P_0^2]}
\ee
where we've used $P_0\equiv P_X(0)$
Here the coupling is seen to explicitly modulate the information flow between $\Xparticle$ and $\Yparticle$. As the coupling $k$ is increased, the logarithmic number of states that the $\Xparticle$ state encodes about the initial $\Yparticle$ state also increases.

Using \equat{power-bound2} we can write $\Ex[Y_0^2]t^2$ as,
\be
\Ex[Y_0^2]t^2=\frac{2m{\Power}_0}{k^2}t-\Ex[X_0^2]t^2
\ee
and substituting this into \equat{XY_approxMutInfo}, along with \equat{energy-bound}, $\Ex[P_0^2]=2m\E_0$, yields,
\be
I(\X_t;\Y_0)\approx\frac{1}{2}\frac{{\Power}_0}{\E_0}t + O(t^2)
\ee
and as before we can compute $\FlowR \equiv\lim_{t\to 0}(I_t\big/t)$, measuring the information flow rate from $\Yparticle$ to $\Xparticle$ in nats/sec, 
\be
\label{eq:R_twoParticle}
\FlowR_{\Yparticle\to\Xparticle} = \frac{1}{2}\left(\frac{{\Power}_0}{\mathcal{E}_0}\right)
\ee
just as we found in \equat{SNR1} and \equat{XF_approxMutInfo}.

\section{Discussion \& Conclusion}
\label{sect:V}
We have argued that the difficulty in defining the information flow between physical constituents amounts to justifying the priors over physical states. Depending on the context, information flow between particles can be infinite, for instance, if a particle state can distinguish between infinitely many environmental states. In other cases, the information exchanged will depend on the choice of priors, which may or may not be justified by an assumption of thermal equilibrium. Outside of such settings, such as far from equilibrium scenarios, what principles can one use to quantify the intuitive sense that information is exchanged between interacting particles?

Our approach to this question is novel in two ways. One is that we frame information exchange between a particle and environment as an additive noise channel, at least over a small time window $t$. We have seen that the previous state of the particle naturally serves as a noise variable in this communication channel. Greater variability over the previous particle state obscures the encoding of the environment in the updated particle state. In contrast, greater variability of environment states contributes to increased information flow as it presents a greater set of states that can be distinguished by the particle. In general an increased space of physical states requires more energy or power. Such variability of particle and environment states is highly context-dependent, however, making exact quantification of information flow difficult. 

This leads to our second contribution. We argue in favor of a saddle-point value for the mutual information as a meaningful measure of information exchange. This saddle-point places a lower bound on the channel capacity between a particle and environment, and can also be interpreted as the maximum information flow in the noisiest setting. This bound allows one to compare information flow in various settings without regard to any particular set of statistics. That is, the bound we construct only requires the energy and power flow between systems, and therefore access to a full distribution over states is unnecessary. Furthermore, this bound is achieved by Gaussian priors and serves as a useful comparison. 

For example, the Brownian buffeting of a particle in thermal diffusion satisfies the assumption we made that $\Ex[(\Dp)^2]=2m{\Power}_0t$, for $\Power_0 = bkT$, with drag coefficient $b$ and bath temperature $T$ ($t\ll b^{-1}$). This is the usual result from the Langevin model, and since $\Dp$ is also assumed to be Gaussian in that context, we can now interpret the ratio,
$$
{\cal R}=\frac{{\Power}_0}{2\E_0} = \frac{bkT}{2(kT_0/2)}=b\left(\frac{T}{T_0}\right)
$$
as the instantaneous information flow rate from a thermal bath to a particle. Here $T_0\equiv 2\E_0/k$ gives the initial temperature of the particle (1-D) before interacting with the bath. As we've discussed, this value is the maximum information flow rate in a maximally noisy setting. In this case, that means that a Gaussian prior with average energy $\E_0$ provides the strongest noise, and the Gaussian buffeting by the bath yields the strongest signal. Thermal buffeting, as in the Langevin model, therefore represents a maximum information flow rate from an environment to a particle. This is not surprising considering the fact that white noise is maximally random, akin to coin flips at each timestep. 

Our result demonstrates a natural relationship between information flow, power and energy in classical particle interactions. The bound ${\Power}_0/2\E_0$ shows that it is the ratio of power and energy that determines the information exchange, and clarifies the intuitive sense that more information can be exchanged in physical degrees of freedom when more energy is allowed to flow. It is interesting that it is the power flow in the zero \textit{average} momentum frame, ${\cal P}_0$, that sets the variance of the environment, and ultimately limits the information flow. $\E_0$ likewise can be computed in a frame where the average momentum is zero, and thus $\Power_0/2\E_0$ can be interpreted as the information flow rate in \textit{that} frame. These quadratic energy constraints originated from the non-relativistic form of the energy. In the zero average momentum frame, interactions with the environment \textit{expand} the phase space of the particle (for small $t>0$). From this perspective, the mutual information computes the extent of this expansion, corresponding to additional degrees of freedom the particle encodes about its environment following the interaction.

We also considered a quadratic potential in a springlike model, and linearity allowed us to compute the mutual information for times beyond a small $t$ window. We
found that the mutual information spikes to infinity periodically. We argued that this corresponds
to one particle's state periodically encoding a single real degree of freedom associated with the
other particle---the infinity here is an artifact of classical physics, where states are points in a continuous phase
space. This result effectively extends classic results on observability (in
regards to the observability of one particle's state given the other) to incorporate an information
theoretic measure quantifying when a state is, at least partially, observable. 

The structure of the saddle-point we consider here is of the form $I_t\approx\E_t\big/2\E_0$ and may suggest a path forward for defining a non-equilibrium temperature for systems arbitrarily far from equilibrium, similar to the work in \cite{Narayanan2012}. In that work, the authors construct a ``thermodynamical temperature'' in analogy with the equilibrium definition of temperature as $T=\Delta E\big/\Delta S$. Along those lines, we can regard the ratio $\E_t/I_t\approx2\E_0$, which is a ratio of average energy flux (from the environment) to the information flow (or negative entropy flow), as a proxy for temperature. In \cite{Narayanan2012}, the authors show that their measure of non-equilibrium temperature aligns with the thermodynamical definition at equilibrium, and thus it would be interesting further work to determine whether our proposed non-equilibrium temperature has the same property.


In future work, we would like to understand whether the results here derive from more general principles regarding information flows in physics models. The inherent length/energy/time scales associated with particle interactions influence information flow and it would be interesting to understand how various course-graining schemes modify the exchange of information between subsystems. How do information flows depend on renormalization, and can one interpret the uncertainty one particle has with regard to its environment more objectively through the lens of the physical scales associated with the particle? A system that is unable to respond sufficiently rapidly to environmental driving cannot encode the environment, for instance, as we saw in the spring model, and the resulting information flows become limited. Initial work along these lines can be found in \cite{Herpich2020}. Broadly understanding how information flows are tied to such physical scales is an interesting direction for future study. This includes the more extreme scales set by quantum and relativistic limits, which were completely ignored in this work.



\section{Appendix}
\subsection{Competitive Optimality of Gaussian Distributions Over Additive Channels}
\label{app:minimax}
It is well known in the communication theory community that under input and noise power constraints ($\Power$ and $\Nzero$,  respectively)  on an additive channel 
\be
\label{eq:+channel}
Y = X + Z
\ee
that the minimax value of mutual information,
\be
I^*(X;Y)
\equiv
\min_{f_{Z}} \max_{f_{X}} I(X;Y) 
=
\max_{f_{X}} \min_{f_{Z}} I(X;Y) 
\ee
is attained when $Z$ and $X$ are Gaussian with $Z \sim {\cal N}(0,\Nzero)$ and $X \sim {\cal N}(0,\Power)$.  This result is also known as competitive optimality of Gaussians for additive channels and figures prominently in the adversarial channel (jamming) literature. \cite{McEliece1983}  That is, appropriate Gaussian distributions on the noise and sender input distributions form a Nash equilibrium from which deviation by the sender will reduce $I(X;Y)$ and from which deviation by the jammer will increase $I(X;Y)$.  We provide a simple proof here for those who may be unfamiliar with the result.

The inequalities in \equat{saddle-point} are given by,
\be
\label{eq:minimaxinequality}
\begin{array}{rccl}
I(X;X+Z^*) & \stackrel{\mbox{(a)}}{\le}  &  I(X^*;X^*+Z^*) & 
\stackrel{\mbox{(b)}}{\le} I(X^*; X^*+Z)\\
\end{array}
\ee
where $Z^*\sim{\cal N}(0,\Nzero)$ and $X^*\sim{\cal N}(0,\Power)$, subject to the constraints,
\be
\label{eq:NPbounds}
\Ex[Z^2]={\Nzero}  \,\,\,\,\, \mbox{and} \,\,\,\,\, \Ex[X^2] = {\Power}
\ee
using $\Nzero\equiv2m\E_0$ and $\Power\equiv2m{\Power}_0t$ as noise and signal power constraints respectively. In terms of the previously defined physical variables we have, 
\be
\begin{array}{ccl}
    Z & = &P_0  \\
    Z^* & = & P_0^* \\
    X & = & \Dp  \\
    X^* & = & \Dp^*.
\end{array}
\ee
In proving the inequalities, we will make use of the following two lemmas. 
\begin{lemma}
    \label{lemma1}
    (max entropy) For a r.v. $W\in\mathbb{R}$ with distribution $f_W$, if $\Ex[W^2]=\E$ then, $$\max_{f_W}h(W)=\frac{1}{2}\log(2\pi e\E)$$which is the entropy of a Gaussian r.v. with variance $\E$.
\end{lemma}
\begin{lemma}
    \label{lemma2} (entropy power inequality) If $X$ and $Y$ are independent r.v.'s and $X'$ and $Y'$ are Gaussian r.v.'s such that $h(X')=h(X)$ and $h(Y')=h(Y)$, then, $$h(X+Y)\ge h(X'+Y')$$
\end{lemma}
Lemma \ref{lemma1} can be found as Theorem 8.6.5 in the classic text \cite{CoverAndThomas}, and Lemma \ref{lemma2} is stated as Theorem 17.8.1 in \cite{CoverAndThomas}.

\underline{The $(a)$ inequality:} $I(X;X+Z^*)\le I(X^*;X^*+Z^*)$. We can write,
\begin{align}
    \begin{split}
        I(X;X+Z^*)&=h(X+Z^*)-h(Z^*) \\
        &=h(X+Z^*)-\frac{1}{2}\log(2\pi e {\Nzero})
    \end{split}
\end{align}
since $X$ and $Z^*$ are assumed independent and the entropy of a Gaussian with variance ${\Nzero}$ is $\frac{1}{2}\log(2\pi e {\Nzero})$. Then note that,
\begin{align}
    \begin{split}
        \Ex[(X+Z^*)^2]&=\Ex[X^2] + \Ex[(Z^*)^2]\\
        &=\sigma_X^2+\overline{X}^2+{\Nzero}\\
        &= {\Power} + {\Nzero}
    \end{split}
\end{align}
since $\Ex[Z^*]=0$ and $\Ex[X^2]=\sigma_X^2+\overline{X}^2$, with $\overline{X}\equiv\Ex[X]$. $\Ex[(Z^*)^2]={\Nzero}$ by definition and $\Ex[X^2]=\Power$ from the constraint in \equat{NPbounds}. And so by Lemma \ref{lemma1}, the maximum entropy of the variable $X+Z^*$, subject to the constraint, $\Ex[(X+Z^*)^2]={\Power}+{\Nzero}$, is given by,
\be
\frac{1}{2}\log(2\pi e({\Power}+{\Nzero}))
\ee
and so we have that,
\begin{align}
    \begin{split}
        I(X;X+Z^*)&=h(X+Z^*)-h(Z^*) \\
        &\le\frac{1}{2}\log(2\pi e({\Power}+{\Nzero}))-\frac{1}{2}\log(2\pi e {\Nzero}) \\
        &=\frac{1}{2}\log \left ( 1 + \frac{{\Power}}{\Nzero} \right )
    \end{split}
\end{align}
which is the mutual information given by $I(X^*;X^*+Z^*)$, concluding the proof of the first inequality. 

\underline{The $(b)$ inequality:} $I(X^*;X^*+Z^*)\le I(X^*;X^*+Z)$. 

We can write, 
\begin{align}
    \begin{split}
        I(X^*;X^*+Z) &= h(X^*+Z) - h(Z) \\
        &\ge h(X^*+Z) - h(Z^*)
    \end{split}
\end{align}
since $X^*$ and $Z$ are assumed independent and the maximum entropy of $Z$, subject to $\Ex[Z^2]=\Nzero$, is given by $h(Z^*)$, the entropy of a Gaussian with variance $\Nzero$. Then, by Lemma \ref{lemma2}, we have that $h(X^*+Z)\ge h(X^*+Z^*)$, and so,
\be
\label{eq:minimax}
I(X^*;X^*+Z) \ge h(X^*+Z^*) - h(Z^*)
\ee
which proves the second inequality. 

This establishes that zero-mean Gaussians for $X$ and $Z$ satisfy a saddle-point condition (sometimes called a minimax condition) for the mutual information.  That is, the minimax mutual information for the additive channel of \equat{+channel} is,
\be
\label{eq:minimaxI}
I^*(X;X+Z)
=
\frac{1}{2}\log\left (
1 + \frac{{\Power}}{{\Nzero}}
\right ).
\ee
The multidimensional case is treated in \cite{Diggavi2001}.

The result of \equat{minimaxI} can be trivially extended to a case where arbitrary means for $\Ex[X] = \Xbar$ and $\Ex[Z] = \Zbar$ are imposed. We define centered variables
$X^\prime = X - \Xbar$ and $Z^\prime = Z - \Zbar$, with suitably adjusted constraints from \equat{NPbounds},
\be
\label{eq:NPboundsnonzeromean}
E[(Z^\prime)^2] = {\Nzero} - \Zbar^2
\,\,\,\,\,\,
\mbox{and}
\,\,\,\,\,\,
E[(X^\prime)^2] = {\Power} - \Xbar^2
\ee
Then, since entropy is invariant under translation, we can recapitulate
\equat{minimaxI} as
\be
\label{eq:minimaxInonzeromean}
I^*(X^\prime;X^\prime+Z^\prime)
=
\frac{1}{2}\log\left (
1 + \frac{{\Power} - \Xbar^2}{{\Nzero} - \Zbar^2}
\right )
\ee
allowing one to incorporate constraints on the mean momentum and force as well.

\subsection{Observability}
\label{app:observability}
In section \ref{sect:IVa} in the main text we stated that the observability matrix has full rank. By this we mean that for system dynamics defined by
\be
\dot{\xv}_t = \Amat \xv_t
\ee
and observable $\X_t$
\be
\X_t = \Cmat \xv_t
\ee
we can always recover $\Y_0$ through observations of $\X_t$ if the observability matrix,
\be
\OImat \equiv 
\left (
\begin{array}{l}
\Cmat \\
\Cmat\Amat\\
\Cmat \Amat^2\\
\Cmat \Amat^3
\end{array} 
\right )
\ee
has full column rank. \cite{ControlTheory2011}  

$\OImat$ having full rank implies the initial state $\Y_0$ can be uniquely determined by a finite sequence of observations $\X_{t_n}$ for distinct times $t_1,t_2,t_3$ and $t_4$, since the total system comprising two one-dimensional particles contains four degrees of freedom, two initial positions and two momenta. Thus, while $\X_t$ at a single time point cannot fully determine $\Y_0$, a sufficient collection of time points can. 

We do find, however, that at certain special times $t$, $\X_t$  \textit{can} be used to determine a single degree of freedom in the pair that comprises the state $\Y_0$. At these times we find that the mutual information is unbounded (spikes), as shown in FIGURE \ref{fig:mutualInfo} in the main text.

\bibliographystyle{apsrev4-2}
\bibliography{refs,McElieceJam}

\begin{thebibliography}{23}%
\makeatletter
\providecommand \@ifxundefined [1]{%
 \@ifx{#1\undefined}
}%
\providecommand \@ifnum [1]{%
 \ifnum #1\expandafter \@firstoftwo
 \else \expandafter \@secondoftwo
 \fi
}%
\providecommand \@ifx [1]{%
 \ifx #1\expandafter \@firstoftwo
 \else \expandafter \@secondoftwo
 \fi
}%
\providecommand \natexlab [1]{#1}%
\providecommand \enquote  [1]{``#1''}%
\providecommand \bibnamefont  [1]{#1}%
\providecommand \bibfnamefont [1]{#1}%
\providecommand \citenamefont [1]{#1}%
\providecommand \href@noop [0]{\@secondoftwo}%
\providecommand \href [0]{\begingroup \@sanitize@url \@href}%
\providecommand \@href[1]{\@@startlink{#1}\@@href}%
\providecommand \@@href[1]{\endgroup#1\@@endlink}%
\providecommand \@sanitize@url [0]{\catcode `\\12\catcode `\$12\catcode `\&12\catcode `\#12\catcode `\^12\catcode `\_12\catcode `\%12\relax}%
\providecommand \@@startlink[1]{}%
\providecommand \@@endlink[0]{}%
\providecommand \url  [0]{\begingroup\@sanitize@url \@url }%
\providecommand \@url [1]{\endgroup\@href {#1}{\urlprefix }}%
\providecommand \urlprefix  [0]{URL }%
\providecommand \Eprint [0]{\href }%
\providecommand \doibase [0]{https://doi.org/}%
\providecommand \selectlanguage [0]{\@gobble}%
\providecommand \bibinfo  [0]{\@secondoftwo}%
\providecommand \bibfield  [0]{\@secondoftwo}%
\providecommand \translation [1]{[#1]}%
\providecommand \BibitemOpen [0]{}%
\providecommand \bibitemStop [0]{}%
\providecommand \bibitemNoStop [0]{.\EOS\space}%
\providecommand \EOS [0]{\spacefactor3000\relax}%
\providecommand \BibitemShut  [1]{\csname bibitem#1\endcsname}%
\let\auto@bib@innerbib\@empty
\bibitem [{\citenamefont {Smoluchowski}(1912)}]{Smoluchowski1912}%
  \BibitemOpen
  \bibfield  {author} {\bibinfo {author} {\bibfnamefont {M.}~\bibnamefont {Smoluchowski}},\ }\href@noop {} {\bibfield  {journal} {\bibinfo  {journal} {Physikalische Zeitschrift}\ }\textbf {\bibinfo {volume} {13}},\ \bibinfo {pages} {1069} (\bibinfo {year} {1912})}\BibitemShut {NoStop}%
\bibitem [{\citenamefont {Feynman}\ \emph {et~al.}(1963)\citenamefont {Feynman}, \citenamefont {Leighton},\ and\ \citenamefont {Sands}}]{Feynman1963}%
  \BibitemOpen
  \bibfield  {author} {\bibinfo {author} {\bibfnamefont {R.~P.}\ \bibnamefont {Feynman}}, \bibinfo {author} {\bibfnamefont {R.~B.}\ \bibnamefont {Leighton}},\ and\ \bibinfo {author} {\bibfnamefont {M.}~\bibnamefont {Sands}},\ }in\ \href@noop {} {\emph {\bibinfo {booktitle} {The Feynman Lectures on Physics, Volume 1}}},\ \bibinfo {editor} {edited by\ \bibinfo {editor} {\bibfnamefont {R.~P.}\ \bibnamefont {Feynman}}, \bibinfo {editor} {\bibfnamefont {R.~B.}\ \bibnamefont {Leighton}},\ and\ \bibinfo {editor} {\bibfnamefont {M.}~\bibnamefont {Sands}}}\ (\bibinfo  {publisher} {Addison-Wesley},\ \bibinfo {address} {Reading, MA},\ \bibinfo {year} {1963})\ Chap.~\bibinfo {chapter} {46}\BibitemShut {NoStop}%
\bibitem [{\citenamefont {Szilard}(1964)}]{Szilard1929}%
  \BibitemOpen
  \bibfield  {author} {\bibinfo {author} {\bibfnamefont {L.}~\bibnamefont {Szilard}},\ }\href {https://doi.org/https://doi.org/10.1002/bs.3830090402} {\bibfield  {journal} {\bibinfo  {journal} {Behavioral Science}\ }\textbf {\bibinfo {volume} {9}},\ \bibinfo {pages} {301} (\bibinfo {year} {1964})}\BibitemShut {NoStop}%
\bibitem [{\citenamefont {Parrando}\ \emph {et~al.}(2015)\citenamefont {Parrando}, \citenamefont {Horowitz},\ and\ \citenamefont {Sagawa}}]{Parrando2015}%
  \BibitemOpen
  \bibfield  {author} {\bibinfo {author} {\bibfnamefont {J.~M.~R.}\ \bibnamefont {Parrando}}, \bibinfo {author} {\bibfnamefont {J.~M.}\ \bibnamefont {Horowitz}},\ and\ \bibinfo {author} {\bibfnamefont {T.}~\bibnamefont {Sagawa}},\ }\href {https://doi.org/10.1038/nphys3230} {\bibfield  {journal} {\bibinfo  {journal} {Nature Physics}\ }\textbf {\bibinfo {volume} {11}} (\bibinfo {year} {2015})}\BibitemShut {NoStop}%
\bibitem [{\citenamefont {Bérut}\ \emph {et~al.}(2012)\citenamefont {Bérut}, \citenamefont {Arakelyan}, \citenamefont {Petrosyan}, \citenamefont {Ciliberto}, \citenamefont {Dillenschneider},\ and\ \citenamefont {Lutz}}]{Landauer-test-1}%
  \BibitemOpen
  \bibfield  {author} {\bibinfo {author} {\bibfnamefont {A.}~\bibnamefont {Bérut}}, \bibinfo {author} {\bibfnamefont {A.}~\bibnamefont {Arakelyan}}, \bibinfo {author} {\bibfnamefont {A.}~\bibnamefont {Petrosyan}}, \bibinfo {author} {\bibfnamefont {S.}~\bibnamefont {Ciliberto}}, \bibinfo {author} {\bibfnamefont {R.}~\bibnamefont {Dillenschneider}},\ and\ \bibinfo {author} {\bibfnamefont {E.}~\bibnamefont {Lutz}},\ }\bibfield  {journal} {\bibinfo  {journal} {Nature}\ }\href {https://doi.org/10.1038/nature10872} {10.1038/nature10872} (\bibinfo {year} {2012})\BibitemShut {NoStop}%
\bibitem [{\citenamefont {Aimet}\ \emph {et~al.}(2025)\citenamefont {Aimet}, \citenamefont {Tajik}, \citenamefont {Tournaire}, \citenamefont {Schüttelkopf}, \citenamefont {Sabino}, \citenamefont {Sotiriadis}, \citenamefont {Guarnieri}, \citenamefont {Schmiedmayer},\ and\ \citenamefont {Eisert}}]{Landauer-test-2}%
  \BibitemOpen
  \bibfield  {author} {\bibinfo {author} {\bibfnamefont {S.}~\bibnamefont {Aimet}}, \bibinfo {author} {\bibfnamefont {M.}~\bibnamefont {Tajik}}, \bibinfo {author} {\bibfnamefont {G.}~\bibnamefont {Tournaire}}, \bibinfo {author} {\bibfnamefont {P.}~\bibnamefont {Schüttelkopf}}, \bibinfo {author} {\bibfnamefont {J.}~\bibnamefont {Sabino}}, \bibinfo {author} {\bibfnamefont {S.}~\bibnamefont {Sotiriadis}}, \bibinfo {author} {\bibfnamefont {G.}~\bibnamefont {Guarnieri}}, \bibinfo {author} {\bibfnamefont {J.}~\bibnamefont {Schmiedmayer}},\ and\ \bibinfo {author} {\bibfnamefont {J.}~\bibnamefont {Eisert}},\ }\bibfield  {journal} {\bibinfo  {journal} {Nature Physics}\ }\href {https://doi.org/10.1038/s41567-025-02930-9} {10.1038/s41567-025-02930-9} (\bibinfo {year} {2025})\BibitemShut {NoStop}%
\bibitem [{\citenamefont {An}\ \emph {et~al.}(2015)\citenamefont {An}, \citenamefont {Zhang}, \citenamefont {Um}, \citenamefont {Lv}, \citenamefont {Lu}, \citenamefont {Zhang}, \citenamefont {Yin}, \citenamefont {Quan},\ and\ \citenamefont {Kim}}]{Jarzynski-test-1}%
  \BibitemOpen
  \bibfield  {author} {\bibinfo {author} {\bibfnamefont {S.}~\bibnamefont {An}}, \bibinfo {author} {\bibfnamefont {J.-N.}\ \bibnamefont {Zhang}}, \bibinfo {author} {\bibfnamefont {M.}~\bibnamefont {Um}}, \bibinfo {author} {\bibfnamefont {D.}~\bibnamefont {Lv}}, \bibinfo {author} {\bibfnamefont {Y.}~\bibnamefont {Lu}}, \bibinfo {author} {\bibfnamefont {J.}~\bibnamefont {Zhang}}, \bibinfo {author} {\bibfnamefont {Z.-Q.}\ \bibnamefont {Yin}}, \bibinfo {author} {\bibfnamefont {H.~T.}\ \bibnamefont {Quan}},\ and\ \bibinfo {author} {\bibfnamefont {K.}~\bibnamefont {Kim}},\ }\href {https://doi.org/10.1038/nphys3197} {\bibfield  {journal} {\bibinfo  {journal} {Nature Physics}\ }\textbf {\bibinfo {volume} {11}},\ \bibinfo {pages} {193} (\bibinfo {year} {2015})}\BibitemShut {NoStop}%
\bibitem [{\citenamefont {Xiong}\ \emph {et~al.}(2018)\citenamefont {Xiong}, \citenamefont {Yan}, \citenamefont {Zhou}, \citenamefont {Rehan}, \citenamefont {Liang}, \citenamefont {Chen}, \citenamefont {Yang}, \citenamefont {Ma}, \citenamefont {Feng},\ and\ \citenamefont {Vedral}}]{Jarzynski-test-2}%
  \BibitemOpen
  \bibfield  {author} {\bibinfo {author} {\bibfnamefont {T.~P.}\ \bibnamefont {Xiong}}, \bibinfo {author} {\bibfnamefont {L.~L.}\ \bibnamefont {Yan}}, \bibinfo {author} {\bibfnamefont {F.}~\bibnamefont {Zhou}}, \bibinfo {author} {\bibfnamefont {K.}~\bibnamefont {Rehan}}, \bibinfo {author} {\bibfnamefont {D.~F.}\ \bibnamefont {Liang}}, \bibinfo {author} {\bibfnamefont {L.}~\bibnamefont {Chen}}, \bibinfo {author} {\bibfnamefont {W.~L.}\ \bibnamefont {Yang}}, \bibinfo {author} {\bibfnamefont {Z.~H.}\ \bibnamefont {Ma}}, \bibinfo {author} {\bibfnamefont {M.}~\bibnamefont {Feng}},\ and\ \bibinfo {author} {\bibfnamefont {V.}~\bibnamefont {Vedral}},\ }\href {https://doi.org/10.1103/PhysRevLett.120.010601} {\bibfield  {journal} {\bibinfo  {journal} {Phys. Rev. Lett.}\ }\textbf {\bibinfo {volume} {120}},\ \bibinfo {pages} {010601} (\bibinfo {year} {2018})}\BibitemShut {NoStop}%
\bibitem [{\citenamefont {Preskill}(2018)}]{Preskill2018}%
  \BibitemOpen
  \bibfield  {author} {\bibinfo {author} {\bibfnamefont {J.}~\bibnamefont {Preskill}},\ }\href {https://doi.org/10.22331/q-2018-08-06-79} {\bibfield  {journal} {\bibinfo  {journal} {Quantum}\ }\textbf {\bibinfo {volume} {2}},\ \bibinfo {pages} {79} (\bibinfo {year} {2018})}\BibitemShut {NoStop}%
\bibitem [{Note1()}]{Note1}%
  \BibitemOpen
  \bibinfo {note} {We assume that the physical laws governing the interaction are fixed and given, so that the information flow is ultimately the result of that structure.}\BibitemShut {Stop}%
\bibitem [{\citenamefont {Schreiber}(2000)}]{Schreiber2000}%
  \BibitemOpen
  \bibfield  {author} {\bibinfo {author} {\bibfnamefont {T.}~\bibnamefont {Schreiber}},\ }\href {https://doi.org/10.1103/PhysRevLett.85.461} {\bibfield  {journal} {\bibinfo  {journal} {Phys. Rev. Lett.}\ }\textbf {\bibinfo {volume} {85}},\ \bibinfo {pages} {461} (\bibinfo {year} {2000})}\BibitemShut {NoStop}%
\bibitem [{Note2()}]{Note2}%
  \BibitemOpen
  \bibinfo {note} {Under the \protect \textit {noisiest} setting, given the constraints, we ask what is the most information that can be transmitted from environment to particle. This question is more meaningful than asking, for instance, what is the most information that can be transmitted in the \protect \textit {least} noisy environment, since that would correspond to zero noise (no variance in the particle prior) and the particle state would exactly encode effect of the environment. The information exchanged would be infinite in this case (for a continuous state space), which is not particularly useful. This highlights the difficulty in defining information flow in a physical setting. The information exchanged depends on the priors over the particle and the environment, but the priors over such variables are context-dependent: The particle could be constrained to be within a box, or in thermal equilibrium at some temperature, etc. Our approach, therefore, is to consider a bound on the mutual information
  over the space of \protect \textit {all} priors such that the mutual information is \protect \textit {maximized} with respect to priors on the environment and \protect \textit {minimized} with respect to priors on the particle state (a maximal signal in a maximal noise setting). We will see that this quantity is finite and hence places a lower bound on the maximum (minimax) information flow between a particle and its environment.}\BibitemShut {Stop}%
\bibitem [{\citenamefont {Rastegar}\ \emph {et~al.}(2012)\citenamefont {Rastegar}, \citenamefont {Roitershtein}, \citenamefont {Roytershteyn},\ and\ \citenamefont {Suh}}]{Rastegar2012}%
  \BibitemOpen
  \bibfield  {author} {\bibinfo {author} {\bibfnamefont {R.}~\bibnamefont {Rastegar}}, \bibinfo {author} {\bibfnamefont {A.}~\bibnamefont {Roitershtein}}, \bibinfo {author} {\bibfnamefont {V.}~\bibnamefont {Roytershteyn}},\ and\ \bibinfo {author} {\bibfnamefont {J.}~\bibnamefont {Suh}},\ }\href {https://doi.org/10.4236/am.2012.312A280} {\bibfield  {journal} {\bibinfo  {journal} {Applied Mathematics}\ }\textbf {\bibinfo {volume} {3}},\ \bibinfo {pages} {2032} (\bibinfo {year} {2012})}\BibitemShut {NoStop}%
\bibitem [{\citenamefont {Seifert}(2012)}]{Seifert2012}%
  \BibitemOpen
  \bibfield  {author} {\bibinfo {author} {\bibfnamefont {U.}~\bibnamefont {Seifert}},\ }\href {https://doi.org/10.1088/0034-4885/75/12/126001} {\bibfield  {journal} {\bibinfo  {journal} {Reports on Progress in Physics}\ }\textbf {\bibinfo {volume} {75}},\ \bibinfo {pages} {126001} (\bibinfo {year} {2012})}\BibitemShut {NoStop}%
\bibitem [{\citenamefont {Cover}\ and\ \citenamefont {Thomas}(2006)}]{CoverAndThomas}%
  \BibitemOpen
  \bibfield  {author} {\bibinfo {author} {\bibfnamefont {T.~M.}\ \bibnamefont {Cover}}\ and\ \bibinfo {author} {\bibfnamefont {J.~A.}\ \bibnamefont {Thomas}},\ }\href@noop {} {\emph {\bibinfo {title} {Elements of Information Theory}}},\ \bibinfo {edition} {2nd}\ ed.\ (\bibinfo  {publisher} {Wiley-Interscience},\ \bibinfo {year} {2006})\BibitemShut {NoStop}%
\bibitem [{Note3()}]{Note3}%
  \BibitemOpen
  \bibinfo {note} {As a point of comparison, the transfer entropy for this example is \protect \textit {infinite}. The TE can be written as $T_{F\to P}=I(P_t;\Delta P_t|P_0)$, which is the mutual information version of equation (4) in \cite {Schreiber2000} with $k=l=1$ and with $n$ and $n+1$ corresponding to the time points $0$ and $t$, respectively. However, in this toy example $I(P_t;\Delta P_t|P_0)=I(P_t;\Delta P_t)+I(P_0;\Delta P_t|P_t)$, since $I(P_0;\Delta P_t)=0$, but the latter quantity explodes, $I(P_0;\Delta P_t|P_t)=\infty $, whereas the former quantity is the finite value we compute, $I(P_t;\Delta P_t)$. The latter quantity, $I(P_0;\Delta P_t|P_t)$, explodes because $P_0$ and $P_t$ together determine the change in the momentum $\Delta P_t$ exactly, and specifying an arbitrary point on the real line, $\Delta P_t\in \protect \mathbb {R}$, requires infinite information. Thus the transfer entropy is not as useful in this context, that is unless one allows for a case like $k=0$ in \cite
  {Schreiber2000}, for which one could say our approach is a \protect \textit {zeroth order} transfer entropy, but that is just the time-lagged mutual information without conditioning.}\BibitemShut {Stop}%
\bibitem [{\citenamefont {{S. Ihara}}(1978)}]{Ihara78}%
  \BibitemOpen
  \bibfield  {author} {\bibinfo {author} {\bibnamefont {{S. Ihara}}},\ }\href@noop {} {\bibfield  {journal} {\bibinfo  {journal} {Information and Control}\ }\textbf {\bibinfo {volume} {37}},\ \bibinfo {pages} {34} (\bibinfo {year} {1978})}\BibitemShut {NoStop}%
\bibitem [{\citenamefont {McEliece}(1983)}]{McEliece1983}%
  \BibitemOpen
  \bibfield  {author} {\bibinfo {author} {\bibfnamefont {R.~J.}\ \bibnamefont {McEliece}},\ }\bibinfo {title} {Communication in the presence of jamming-an information-theoretic approach},\ in\ \href {https://doi.org/10.1007/978-3-7091-2640-0_8} {\emph {\bibinfo {booktitle} {Secure Digital Communications}}},\ \bibinfo {editor} {edited by\ \bibinfo {editor} {\bibfnamefont {G.}~\bibnamefont {Longo}}}\ (\bibinfo  {publisher} {Springer Vienna},\ \bibinfo {address} {Vienna},\ \bibinfo {year} {1983})\ pp.\ \bibinfo {pages} {127--166}\BibitemShut {NoStop}%
\bibitem [{\citenamefont {J.M.~Borden}(1985)}]{BordenMasonMcEliece85}%
  \BibitemOpen
  \bibfield  {author} {\bibinfo {author} {\bibfnamefont {R.~M.}\ \bibnamefont {J.M.~Borden}, \bibfnamefont {D.M.~Mason}},\ }\href@noop {} {\bibfield  {journal} {\bibinfo  {journal} {{SIAM J. Control and Optimization}}\ }\textbf {\bibinfo {volume} {23}} (\bibinfo {year} {1985})}\BibitemShut {NoStop}%
\bibitem [{\citenamefont {Diggavi}\ and\ \citenamefont {Cover}(2001)}]{Diggavi2001}%
  \BibitemOpen
  \bibfield  {author} {\bibinfo {author} {\bibfnamefont {S.}~\bibnamefont {Diggavi}}\ and\ \bibinfo {author} {\bibfnamefont {T.}~\bibnamefont {Cover}},\ }\href {https://doi.org/10.1109/18.959289} {\bibfield  {journal} {\bibinfo  {journal} {IEEE Transactions on Information Theory}\ }\textbf {\bibinfo {volume} {47}},\ \bibinfo {pages} {3072} (\bibinfo {year} {2001})}\BibitemShut {NoStop}%
\bibitem [{\citenamefont {Herpich}\ \emph {et~al.}(2020)\citenamefont {Herpich}, \citenamefont {Shayanfard},\ and\ \citenamefont {Esposito}}]{Herpich2020}%
  \BibitemOpen
  \bibfield  {author} {\bibinfo {author} {\bibfnamefont {T.}~\bibnamefont {Herpich}}, \bibinfo {author} {\bibfnamefont {K.}~\bibnamefont {Shayanfard}},\ and\ \bibinfo {author} {\bibfnamefont {M.}~\bibnamefont {Esposito}},\ }\href {https://doi.org/10.1103/PhysRevE.101.022116} {\bibfield  {journal} {\bibinfo  {journal} {Phys. Rev. E}\ }\textbf {\bibinfo {volume} {101}},\ \bibinfo {pages} {022116} (\bibinfo {year} {2020})}\BibitemShut {NoStop}%
\bibitem [{\citenamefont {Dahleh}\ \emph {et~al.}(2011)\citenamefont {Dahleh}, \citenamefont {Dahleh},\ and\ \citenamefont {Verghese}}]{ControlTheory2011}%
  \BibitemOpen
  \bibfield  {author} {\bibinfo {author} {\bibfnamefont {M.}~\bibnamefont {Dahleh}}, \bibinfo {author} {\bibfnamefont {M.~A.}\ \bibnamefont {Dahleh}},\ and\ \bibinfo {author} {\bibfnamefont {G.}~\bibnamefont {Verghese}},\ }\href@noop {} {\bibinfo {title} {Lectures on dynamic systems and control}},\ \bibinfo {howpublished} {\href{https://ocw.mit.edu/courses/6-241j-dynamic-systems-and-control-spring-2011/2f03f88e1a714f3ccdb0b7b3f05a2c55_MIT6_241JS11_chap24.pdf}} (\bibinfo {year} {2011}),\ \bibinfo {note} {{Lecture notes for 6.241J Dynamic Systems and Control (Spring 2011)}}\BibitemShut {NoStop}%
\bibitem [{\citenamefont {Narayanan}\ and\ \citenamefont {Srinivasa}(2012)}]{Narayanan2012}%
  \BibitemOpen
  \bibfield  {author} {\bibinfo {author} {\bibfnamefont {K.~R.}\ \bibnamefont {Narayanan}}\ and\ \bibinfo {author} {\bibfnamefont {A.~R.}\ \bibnamefont {Srinivasa}},\ }\href {https://doi.org/10.1103/PhysRevE.85.031151} {\bibfield  {journal} {\bibinfo  {journal} {Physical Review E}\ }\textbf {\bibinfo {volume} {85}},\ \bibinfo {pages} {031151} (\bibinfo {year} {2012})}\BibitemShut {NoStop}%
\end{thebibliography}%

\end{document}